\documentclass[aps,prb,twocolumn,showpacs]{revtex4}
\usepackage{graphicx,color,dcolumn,bm,float,longtable,mathptmx}
\usepackage{amsfonts,amsmath,amssymb}

\begin{document} 
\title{Jahn-Teller driven perpendicualr magnetocrystalline anisotropy in metastable Ru}
\author{Dorj Odkhuu$^{1,2}$}
\author{S. H. Rhim$^{1,3}$}
\email[Email address: ]{sonny@ulsan.ac.kr}
\author{Noejung Park$^4$}
\author{Kohji Nakamura$^5$}
\author{Soon-Cheol Hong$^1$}
\email[Email address: ]{schong@ulsan.ac.kr}
\affiliation{
 $^1$ Department of Physics and Energy Harvest Storage Research Center, University of Ulsan, Ulsan,Korea\\
 $^2$ Department of Physics, Incheon National University, Incheon, Korea
 $^3$ Department of Physics and Astronomy, Northwestern University, Evanston,IL,60208\\
 $^4$ Department of Physics, UNIST, Ulsan, Korea\\
 $^5$Department of Physics Engineering, Mie University,Tsu, Mie, 514-8507, Japan
} 
\date{\today}

\begin{abstract}
  A new metastable phase of the body-centered-tetragonal ruthenium ({\em bct}--Ru) 
  is identified to exhibit a large perpendicular magnetocrystalline anisotropy (PMCA),
  whose energy, $E_{MCA}$, is as large as 150 $\mu$eV/atom, two orders of magnitude greater
  than those of 3$d$ magnetic metals.
  Further investigation over the range of tetragonal distortion suggests that 
  the appearance of the magnetism in the {\em bct}--Ru is governed by the Jahn-Teller spit $e_g$ orbitals.
  Moreover, from band analysis, 
  MCA is mainly determined by an interplay between two $e_g$ states,
  $d_{x^2-y^2}$ and $d_{z^2}$ states, as a result of level reversal associated with tetragonal distortion.
\end{abstract}
\maketitle  

Physics phenomena originated from spin-orbit interaction,
such as magneto-crystalline anisotropy (MCA),
Rashba-type interactions, or topological insulator,
have attracted huge attention
for its intriguing physics as well as great potential
for spintronics applications.\cite{Hasan:rmp2010,Qi:rmp2011,wolf:sci01,zuticc:rmp04,brataas12:nmat}
In particular, MCA,
where one particular direction of the magnetization is energetically preferred,
offers opportunities in spintronics such as
magnetic random access memory (MRAM),
spin-transfer torque (STT), magneto-optics, and to list a few.
With advances of fabrication techniques in recent years,
search for materials with large MCA,
more preferrably perpendicular MCA (PMCA),
has been very intensive.

In particular, 
ferromagnetic films that can provide perpendicular MCA (PMCA)
are indispensable constituents
in STT memory that utilizes spin-polarized tunneling current
to switch magnetization.\cite{sun00:prb}
For practical operation of high-density memory bits,
two criteria have to be satisfied for practical usage 
of high-density magnetic storage
- low switching current ($I_{SW}$) and thermal stability. 
Small volume is favored to lower $I_{SW}$,
but is detrimental for the thermal stability.
However, 
the shortcoming of small volume 
can be compensated by large MCA
while still retaining the thermal stability.
On the other hand, low magnetization will offer advantage
to reduce stray field in real devices.
Therefore, exploration for materials with high anisotropy and small magnetization
would be one favorable direction
to minimize $I_{SW}$ and at the same time to maximize the thermal stability.

\begin{figure*}[htpb]
  \centering
  \includegraphics[width=\columnwidth]{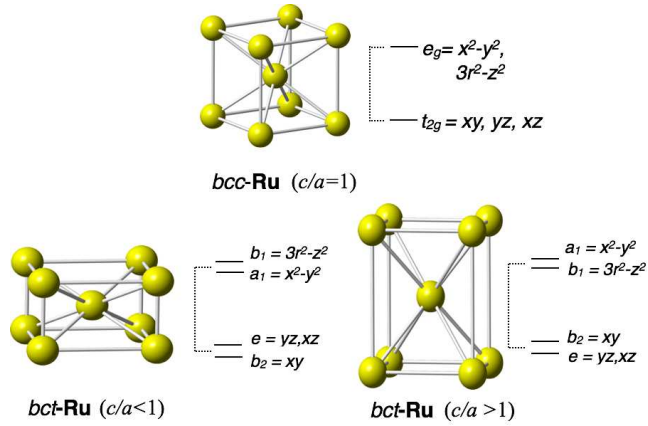}
  \caption{(color online) Schematic presentation of the Jahn-Teller splitting of {\em d} electrons.
    In the cubic symmetry, such as {\em bcc} ($c/a=1$), {\em d} orbital splits into
    doublet ($e_g$) and triplet ($t_{2g}$).
    Tetragonal distortion further splits
    $e_g$ into $a_1$ and $b_1$; $t_{2g}$ into a singlet $b_2$ and a doublet $e$,
    where their relative order is shown depending on $c/a$ is greater or smaller than unity.
}
  \label{fig:1}
\end{figure*}
In the framework of perturbation theory\cite{wang93:prb},
$E_{MCA}$ is determined by the spin-orbit interaction
between occupied and unoccupied states as,
\begin{equation}
  \label{eq:EMCA_pert}
  E^{\sigma\sigma'}_{MCA}\approx \xi^2 \sum_{o,u} 
  \frac{|\langle o^{\sigma} | \ell_{Z} | u^{\sigma'} \rangle |^2
      - |\langle o^{\sigma} | \ell_{X} | u^{\sigma'} \rangle |^2}
  { \epsilon_{u,\sigma'} - \epsilon_{o,\sigma} },
\end{equation}
where $o^{\sigma}$ ($u^{\sigma'}$) and $\epsilon_{o,\sigma}$ ($\epsilon_{u,\sigma'}$) represent
eigenstates and eigenvalues of occupied (unoccupied) for each spin state,
$\sigma,\sigma'=\uparrow,\downarrow$, respectively;
$\xi$ is the strength of spin-orbit coupling (SOC).

As the electronic structure of magnetic materials with non-negigible MCA 
is mainly dominated by $d$ electrons, 
it would be worthwhile to see
how energy levels of $d$ orbitals evolve in different crystal symmetry,
as illustrated in Fig.~\ref{fig:1}.
For the {\em bcc}-Ru with $c/a=1$,
the cubic symmetry splits five {\em d} orbitals
into doublet ($e_g$) and triplet ($t_{2g}$).
When the lattice changes from high-symmetric body-centered
to tetragonal with lower symmetry,
additional Jahn-Teller splitting may offer more
freedom to provide more energy differences in Eq.~(\ref{eq:EMCA_pert}).
More specifically, {\em d} electrons in the {\em bcc} structure 
split into doublet ($e_g$) and triplet ($t_{2g}$).
The tetragonal distortion further splits
these $e_g$ and $t_{2g}$ levels
into two irreducible representations: $e_g$ into two singlets
$a_1$ ($d_{z^2}$) and $b_1$ ( $d_{x^2-y^2}$); 
$t_{2g}$ into a singlet $b_2$ ($d_{xy}$) and a doublet $e$ ($d_{yz,xz}$),
where their relative order is determined by $c/a$, either larger or smaller than unity.

Metals with $4d$ and $5d$ valence electrons possess inherently larger SOC
than conventional $3d$ metals.
Search for magnetism in these transition metals have a long history.
The fact that Pd and Pt barely miss the Stoner criteria to become ferromagnetic (FM)
has incurred enormous efforts to realize magnetism 
in several multilayers and interfaces of 4$d$ metals
by adjusting volumes or lattice constants,
thereby increased density of states (DOS) at the Fermi level ($E_F$), $N(E_F)$,
due to narrowed bandwidth, would meet the Stoner criteria. 
Hence, $4d$ and $5d$ metals with large SOC as well as magnetism
would be favorable candidate to realize large MCA.

Previous theoretical study suggested that ferromagnetism in Ru is feasible
in body-centered-cubic ({\em bcc}) structure when lattice is expanded by 5\%.\cite{kobayashi94:jpc}
Other studies predicted that magnetism can occur in Rh and Pd with volume changes.
\cite{moruzzi89:prb,chen89:prb}
However, those theoretically proposed magnetism associated with volume changes in $4d$ metals
have not been fully confirmed experimentally.
Nevertheless, with remarkable advances in recent fabrication techniques,
various types of lattices 
are now accessible with diverse choice of substrates.
In particular, the {\em bct}--Ru film has been successfully fabricated on the Mo (110) substrate, 
whose lattice constants are $a$=3.24~\AA~and $c/a$=0.83
as identified by X-ray electron diffraction.\cite{shiiki97:jjap} 
Later, theoretical calculation argued that
magnetism can exist in the {\em bct}--Ru for $c/a=0.84$
with moment of 0.4 $\mu_B$/atom.\cite{watanabe00:jmmm}

In this paper, we present that 
in a newly identified metastable phase of the {\em bct}--Ru,
$E_{MCA}$ can be as large as 150 $\mu$eV/atom,
two orders of magnitude greater than those in $3d$ magnetic metals.
The magnetic instability driven
by this tetragonal distortion is discussed in connection with the Stoner criteria.
Furthermore, we show that magnetism as well as MCA
are governed mainly by the Jahn-Teller split $e_g$ orbitals.

\begin{table*}[htpb]
  \centering
  \caption{Calculated equilibrium lattice parameters, $a$ and $c/a$ (in \AA),
    and total energy difference $\Delta E$ (in eV/atom)
    of {\em hcp}-, {\em fcc},- {\em bcc}-, and {\em bct}-Ru
    with respect to the total energy of {\em hcp} structure.
    Experimental and previous theoretical results are also given for comparison. }
  \label{tab:1}
\begin{ruledtabular}
  \begin{tabular}{cccccccccc}
& \multicolumn{2}{c}{{\em hcp}} &\multicolumn{2}{c}{{\em fcc}} &\multicolumn{2}{c}{{\em bcc}} & \multicolumn{3}{c}{{\em bct}} \\
\hline
& Present&Experiment\footnotemark[1]&Present&Previous\footnotemark[2]&Present&Previous   & Present & Previous\footnotemark[2] & Experiment\footnotemark[1]\\    
\hline
 $a$   & 2.70    & 2.70   &   3.84    & 3.84    &  3.07    & 3.06               & 3.25    & 3.25     &  3.24    \\
 $c/a$ & 1.58    & 1.58   &   1.09    & 1.00    &  1.00    & 1.00               & 0.84    & 0.83     &  0.83    \\
$\Delta E$ & 0.0 &         &   0.07    & 0.13    &  0.56    & 0.65               & 0.48    & 0.55     &  -       \\
  \end{tabular}
\end{ruledtabular}
\footnotetext[1]{Shiiki {\em et al.}  \cite{shiiki97:jjap} }
\footnotetext[2]{Watanabe {\em et al.} \cite{watanabe00:jmmm} }
\end{table*}

\begin{figure}[b]
  \centering
  \includegraphics[width=\columnwidth]{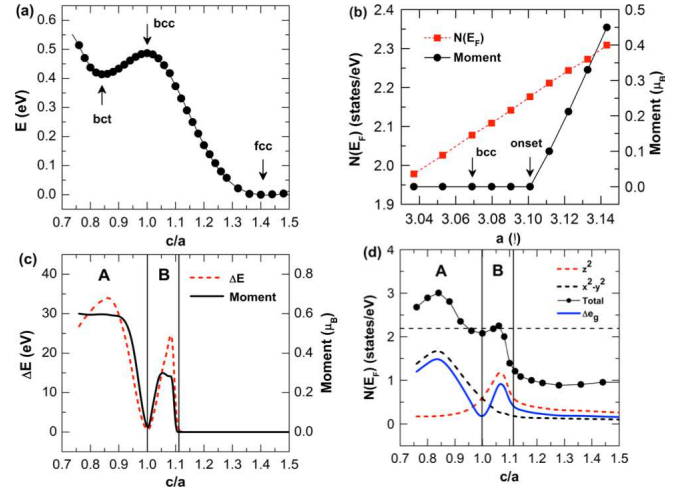}
  \caption{(color online)
    (a) Total energy with respect to {\em fcc} structure ($c/a$=1.41) of
    non-magnetic {\em bct}--Ru upon the tetragonal distortion ($c/a$)
    in fixed volume of the {\em bcc} structure.
    The equilibrium $c/a$ for {\em bct}, {\em bcc}, and {\em fcc}
    are denoted.
    (b) $N(E_F)$ of non–-spin-–polarized calculations (red squares),
    and magnetic moment of the {\em bcc}--Ru as a function of the uniform lattice constant $a$
    (black circles).
    The arrow denotes the equilibrium lattice constant of {\em bcc}−-Ru.
    (c) Energy difference $\Delta E = E_{NM}- E_{FM}$ (red dotted line),
    magnetic moments (black solid line) as function of $c/a$.
    The tetragonal distortion is classified into two regions,
    $A$ and $B$, by $c/a<1$ or $>1$.
    (d) $N(E_F)$ of NM {\em bct}--Ru as function of $c/a$.
    Total $N(E_F)$, those from $d_{z^2}$, $d_{x^2-y^2}$,
    and the absolute value of the difference of the two $e_g$ orbitals,
    denoted as $\Delta e_{g}$,
    are shown in black solid circles, black dotted line, red dashed line, and blue solid line, respectively.
 }
  \label{fig:2}
\end{figure}
Density functional calculations were performed using
the highly precise full-potential linearized augmented plane wave (FLAPW) method.\cite{wimmer:81}
For the exchange--correlation potential, generalized gradient approximation (GGA) was employed
as parametrized by Perdew, Burke and Ernzerhof (PBE).\cite{perdew96:prl}
Energy cutoffs of 16 and 256 Ry were used for wave function expansions and potential representations.
Charge densities and potential inside muffin-tin (MT) spheres
were expanded with lattice harmonics $\ell \leq 8 $
with MT radius of 2.4 a.u.
To obtain reliable values of MCA energy ($E_{MCA}$), 
calculations with high precision is indispensable.
A 40$\times$40$\times$40 mesh in the irreducible Brillouin zone wedge 
is used for {\em k} point summation.
A self-−consistent criteria of 1.0$\times$10$^{−-5}$ $e/(a.u.)^3$ was imposed for calculations,
where convergence with respect to the numbers of basis functions
and {\em k} points was also seriously checked.\cite{mokrousov06:prl,hong08:jkps}
For the calculation of $E_{MCA}$,
torque method\cite{wang93:prb,wang96:prb_torque} was employed to reduce computational costs,
whose validity and accuracy have been proved
in conventional FM materials.\cite{wu99:jmmm,odkhuu10:jap,zhang10:prb82,odkhuu11:apl,odkhuu12:jap,odkhuu13:prb}

Equilibrium lattice constants of hexagonal-closed-packed ({\em hcp})-,
face-center-cubic ({\em fcc})-, and {\em bcc}-Ru
are summarized in Table~\ref{tab:1}, which are in good agreement
with experiments,\cite{shiiki97:jjap,lide02:handbook} and previous work.\cite{watanabe00:jmmm}
The {\em hcp} structure is the most stable phase, as Ru crystallizes in {\em hcp}. 
However, the energy difference between {\em hcp} and {\em fcc}, 0.07 eV/atom, is very small,
which reflects the feature of closed packed structures of the two but with different stacking sequences.
In Fig.~\ref{fig:2}(a) total energy of non-magnetic (NM) {\em bct}--Ru 
as a function of tetragonal distortion ($c/a$) 
is plotted for the fixed volume of the equilibrium {\em bcc-}structure.
Our result reproduces that by Watanabe {\em et al.}\cite{watanabe00:jmmm}:
There is a global minimum at $c/a=1.41$ corresponding to the {\em fcc} structure.
There are two other extrema, a local maximum and minimum 
at $c/a=1$ and $c/a=0.84$, respectively.
In particular, the local minimum at $c/a=0.84$ suggests
the existence of metastable phase as discussed in Ref.\cite{watanabe00:jmmm}. 
Further calculations of total energy of the {\em bct} structure as function of 
both $a$ and $c/a$ confirms that the local minimum is at $a=3.25~\AA$ and $c/a=0.84$,
consistent with the fixed volume calculation of the {\em bcc} structure.

In Fig.~\ref{fig:2}(b),
$N(E_F)$ of non-spin-polarized and magnetic moment of spin-polarized calculation are plotted
as function of lattice constant.
The onset of magnetism in the {\em bcc} phase occurs at $a=$3.10~\AA,
which corresponds to 1.1\% expansion of lattice constant, {\em or}
3.3\% expansion of volume, as consistent with Ref.\cite{kobayashi94:jpc}.
In order for the magnetic instability in the {\em bcc} phase to satisfy
the Stoner criteria, $I\cdot N(E_F) \geq 1$, and
from the fact that the Stoner factor $I$ of a particular atom
does not differ substantially in different crystal structures,
we estimate $I=0.46$ eV for Ru from $N(E_F)=2.18$ eV$^{-1}$.

On the other hand,
as shown in Fig.~\ref{fig:2}(c),
the energy difference between NM and FM states
($\Delta E = E_{NM}- E_{FM}$) and magnetic moment reveal almost the same trends as $c/a$ changes.
$\Delta E$ of the {\em bcc}- and {\em fcc}-phases are negligibly small,
thus both phases are non-magnetic.
When $c/a<1.1$ but $c/a\ne1$, the {\em bct}--Ru is magnetic ($\Delta E>0$), 
whereas $c/a>1.1$, it is non-magnetic.
In particular, $c/a=0.84$ gives $\Delta E$=35 meV/atom
with magnetic moment as high as 0.6 $\mu_B$,
larger than 0.40 $\mu_B$ by Ref.\cite{watanabe00:jmmm}.
Interestingly, the magnetic moment of the {\em bct}--Ru exhibits a re-entrance behavior for $c/a>1$,
as predicted by Sch\"onecker {\em et al.}\cite{schoenecker12:prb}.
In region $A$ ($c/a < 1 $),
magnetic moment decreases as $c/a$ increases,
whereas magnetism reappears when $c/a$ just passes unity,
which eventually vanishes for $c/a > 1.1$. 

\begin{figure}[htpb]
  \centering
  \includegraphics[width=\columnwidth]{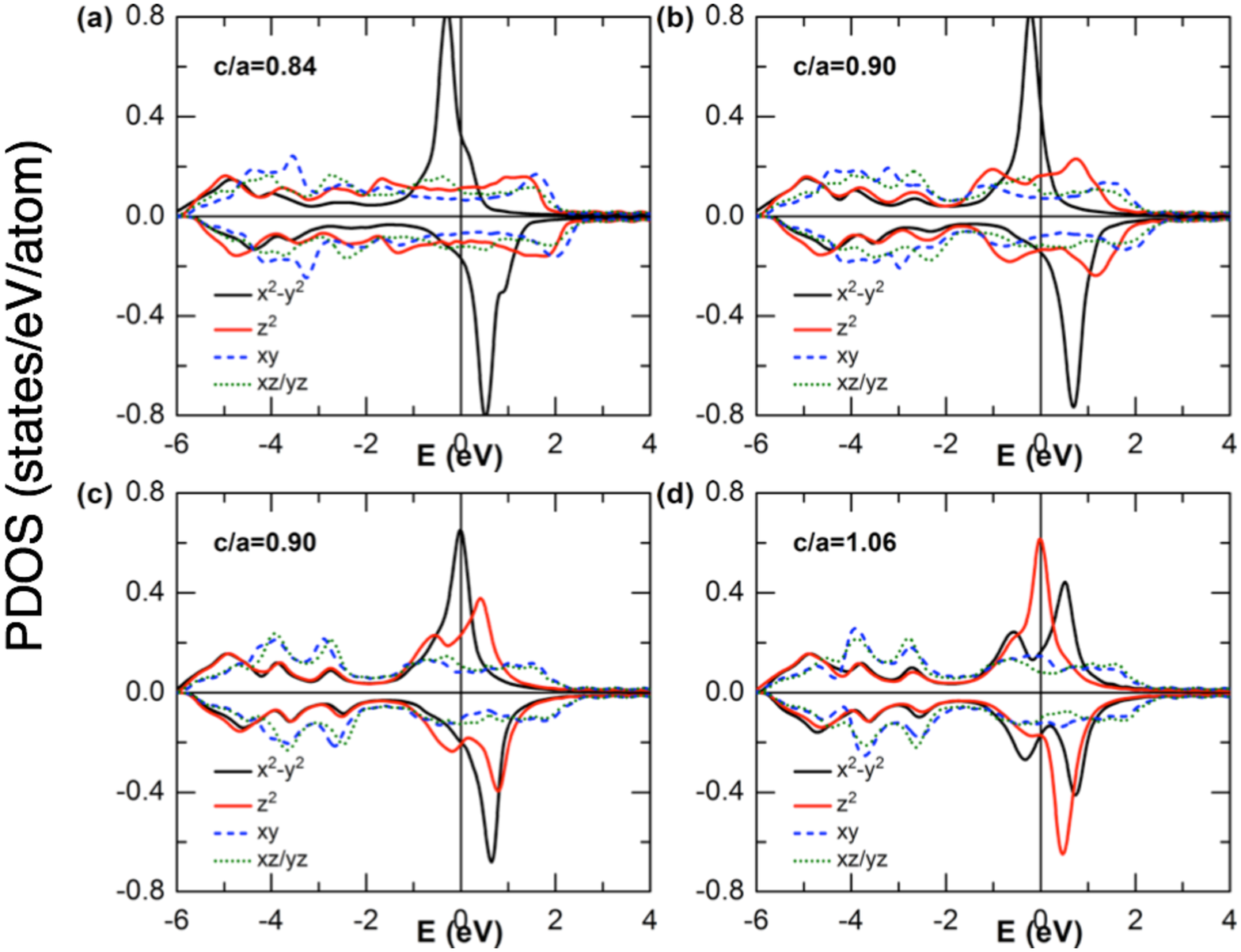}
  \caption{(color online)
    Orbital-−decomposed DOS of {\em d}-–orbital for spin–polarized calculations of {\em bct}-–Ru at
     $c/a$= (a) 0.84, (b) 0.90, (c) 0.96, and (d) 1.06, respectively.
     The $d$ orbital states are shown in different colors: 
     red ($d_{z^2})$, black ($d_{x^2-y^2}$), blue ($d_{xy}$), and green ($d_{xz,yz}$), respectively.
}
\label{fig:3}
\end{figure}

Total DOS and those from $e_g$ orbitals at $E_F$ as function of $c/a$
are plotted in Fig.~\ref{fig:2}(d) for the NM {\em bct}--Ru.
Most contributions come from the Jahn-Teller split $e_g$ orbitals,
whose difference in DOS is also plotted:
It resembles magnetic moment shown in Fig.~\ref{fig:2}(c).
Moreover, among the Jahn-Teller split $e_g$ orbitals,
$d_{x^2-y^2}$ ( $d_{z^2}$ ) dominates the other for $c/a<1$ ($c/a>1$).

Partial DOS (PDOS) of $d$ orbitals are shown in Fig.~\ref{fig:3}
for the spin-polarized cases,
where the trivial $c/a=1$ is omitted.
Prominent peaks at $c/a=0.84$ are mainly from $d_{x^2-y^2}$ states with 
occupied (unoccupied) peaks in majority (minority) spin bands,
while peaks in $d_{z^2}$ states evolve as $c/a$ increases.
Contributions from $t_{2g}$ states are rather featureless.

For simplicity, we assign the energy difference of peaks in $e_g$ states,
$d_{x^2-y^2}$ for $c/a < 1 $ and $d_{z^2}$ for $c/a>1$, respectively, 
as the exchange-splitting.
Then, as $c/a$ increases, the exchange-splittings 
are 1.02, 1.05, 0.80, and 0.66 eV for $c/a=$ 0.84, 0.90, 0.96, and 1.06, respectively,
which qualitatively reflects magnetism of the {\em bct}--Ru.
From this,
the exchange-splitting is mainly determined by one of the Jahn-Teller split $e_{g}$ orbitals.

\begin{figure}[t]
\centering
\includegraphics[width=\columnwidth]{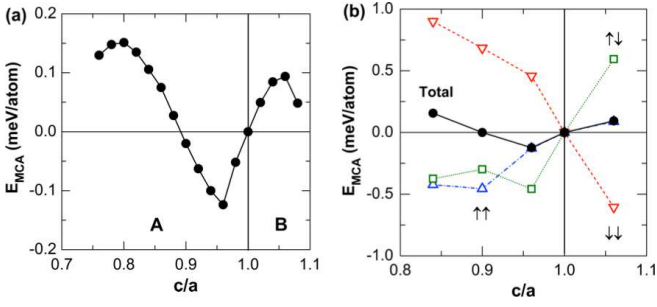}
\caption{(color online)
 (a) MCA energy dependence on $c/a$ for {\em bct}--Ru, where A−B are defined as in Fig.~\ref{fig:2}(b).
 (b) Spin-channel decomposed and total $E_{MCA}$ of
 {\em bct}-Ru for various $c/a$.
 Black circles denote total MCA.
 Upper (lower) triangles denote $\uparrow\uparrow$ ($\downarrow\downarrow)$-channel,
 squares denote $\uparrow\downarrow$-channel. }
\label{fig:4}
\end{figure}

In addition to magnetism, 
the {\em bct}--Ru exhibits large MCA.
The angle-dependent total energy in a tetragonal symmetry is expressed in the most general form,
$E_{tot}(\theta,\varphi) = E_{0}+k_1 \sin^{2}\theta +k_2\sin^{4}\theta +k_3\sin^{4}\theta \cos4\varphi,$ 
where $\theta$ and $\varphi$ are polar and azimuthal angles, respectively,
and $k_1=100$, $k_2=-1$, and $k_3 \ll 1~\mu$eV.
The small value of $k_3$ indicates negligible $\varphi$ dependence. 
$E_{MCA}=E_{tot}(\theta=90^{\circ})-E_{tot}(\theta=0^{\circ}) $
as function of the tetragonal distortion $c/a$ is shown in Fig.~\ref{fig:4}(a).
$E_{MCA}$=150 $\mu$eV/atom at $c/a=0.80$,
and $E_{MCA}$=100 $\mu$eV/atom for the local minimum ($c/a=0.84$),
which are two orders of magnitude greater than conventional 3$d$ magnetic metals.
As the strength of the tetragonal distortion changes,
$E_{MCA}$ changes not only in magnitude but also in sign.
In region $A$, $E_{MCA}$ becomes negative near $c/a\approx 0.9$
and reaches $-100$~ $\mu$eV/atom around $c/a=0.96$,
whereas in region $B$, $E_{MCA}>0$ : PMCA is restored.
Hence, the strength of the tetragonal distortion, $c/a$,
influences magnetic moments as well as $E_{MCA}$.

$E_{MCA}$ is decomposed into different spin-channels following Eq.~(\ref{eq:EMCA_pert}),
as shown in Fig.~\ref{fig:4}(b)
for the {\em bct}--Ru with $c/a$= 0.84, 0.90, 0.96 and 1.06, respectively.
For $\sigma\sigma'=\uparrow\uparrow$ or $\downarrow\downarrow$,
positive (negative) contribution to $E_{MCA}$ is determined by
the SOC interaction between occupied and unoccupied states
with the same (different by one) magnetic quantum number ($m$) through
the $\ell_{Z}$ ($\ell_{X}$) operator.
For $\sigma \sigma'=\uparrow\downarrow$, Eq.~(\ref{eq:EMCA_pert}) has opposite sign,
so positive (negative) contribution come from the $\ell_{X}$ ($\ell_{Z}$) coupling.

From the spin-channel decomposition of $E_{MCA}$,
one notes that there is no dominant spin-channel.
This feature differs from the 3$d$ transition metals,
where particular spin-channel, i.e. the $\downarrow\downarrow$-channel, dominantly
contributes to positive value through  
the SOC matrix $\langle x^2-y^2 | \ell_{Z}| xy\rangle$
with negligible ones from $\ell_{X}$ matrices.\cite{wang93:prb,hotta13:prl}
When $c/a=0.84$, the $\downarrow\downarrow$-channel gives the largest contribution,
while those from other channels are smaller than half of the $\downarrow\downarrow$-channel
with opposite signs.
As $c/a$ increases,
the $\downarrow\downarrow$-channel contribution is reduced,
which turns negative for $c/a>1$.
MCA almost vanishes for $c/a=0.90$ and becomes negative for $c/a=0.96$.
On the other hand,
for $c/a=1.06$, 
the $\uparrow\downarrow$- and $\downarrow\downarrow$-channels contribute
almost the same magnitudes with opposite signs,
so just the $\uparrow\uparrow$-channel contribution remains.

\begin{figure}
  \centering
  \includegraphics[width=\columnwidth]{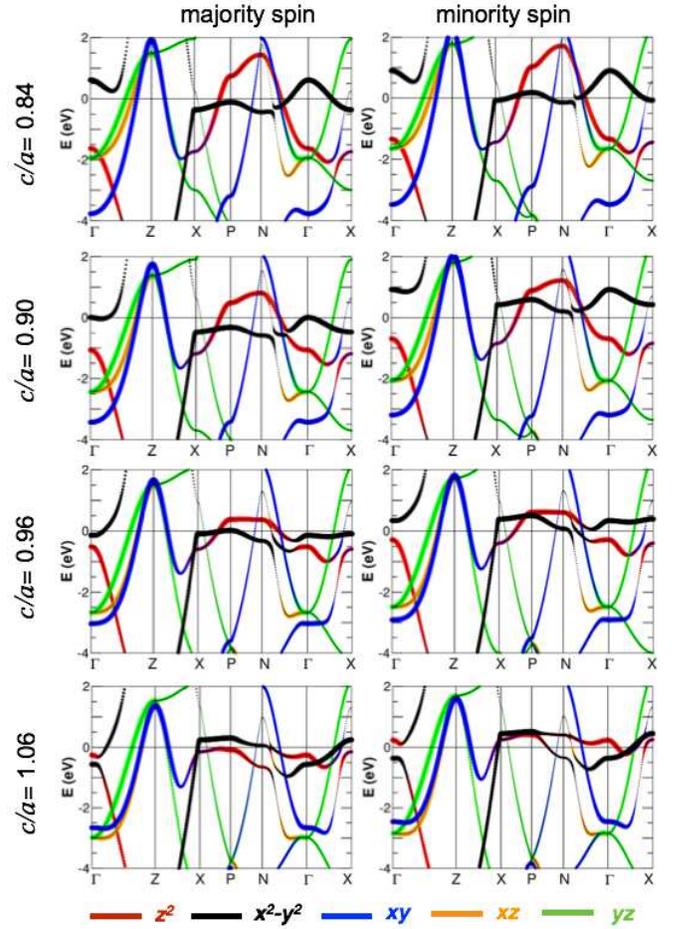}
  \caption{(color online) Band structures of {\em bct}--Ru for $c/a$=0.84, 0.90, 0.96, and 1.06
    for majority and minority spin states.
    $d$ orbital states are shown in different colors: red ($d_{z^2}$),
    black ($d_{x^2-y^2}$), blue ($d_{xy}$),
    orange ($d_{xz}$), and green ($d_{yz}$), respectively.
}
  \label{fig:5}
\end{figure}

To obtain more insights,
band structure is plotted in Fig.~\ref{fig:5}
with $d$ orbital projection,
where size of symbols is proportional to their weights.
All bands along the $\Gamma$-$Z$-$X$ are highly dispersive,
whereas those along the $X$-$P$-$N$-$\Gamma$-$X$ are less dispersive
with rather flat feature from $d_{x^2-y^2}$ and $d_{z^2}$ states.
Level reversals between $e_g$ states, $d_{x^2-y^2}$ and $d_{z^2}$,
are well manifested,
while $t_{2g}$ states are relatively rigid with respect to tetragonal distortion.
It is a formidable task to identify
the role of each individual SOC matrix for each $c/a$.
However, from the spin-channel decomposed MCA [Fig.~\ref{fig:4}(b)],
each spin-channel changes its sign when $c/a$ becomes greater than unity,
where the level reversal occurs between $d_{x^2-y^2}$ and $d_{z^2}$.

For a simple analysis,
we express the $\downarrow\downarrow$-channel as
\begin{equation}
  \label{eq:MCA_}
  E(\downarrow\downarrow) = 
  \frac{|\langle x^2-y^2|\ell_{Z}|xy\rangle|^2}{\epsilon_{x^2-y^2} - \epsilon_{xy}}
  - \frac{|\langle x^2-y^2|\ell_{X}|xz\rangle|^2}{\epsilon_{x^2-y^2} - \epsilon_{xz}}
  - \frac{|\langle z^2|\ell_{X}|xz\rangle|^2}{\epsilon_{z^2} - \epsilon_{xz}}.
\end{equation}
We focus along the $X$-$P$-$N$-$\Gamma$-$X$,
where $e_g$ are unoccupied.
The $\langle yz|\ell_{Z}| xz\rangle$ contributions are neglected
due to the rigidity of $t_{2g}$ states
as well as their small contribution to $E_{MCA}$ owing to large energy denominator.
From the fact that $E(\downarrow\downarrow)>0$ when $c/a=0.84$ with the largest value,
we can infer that the first term in Eq.~(\ref{eq:MCA_}) should be larger than the other two,
where the largest occur along the $P$-$N$.
[See Supplementary Information (SI) for the {\em k}-resolved MCA analysis.]

As $c/a$ increases but $c/a<1$, 
the empty $d_{z^2}$ band moves downward
while the empty $d_{x^2-y^2}$ band goes upward with respect to $E_F$.
As a result, the third term is enhanced due to smaller energy denominator.
Hence, $E(\downarrow\downarrow)$ decreases but remains positive.
When $c/a>1$, however, the level reversal between $e_{g}$ states pushes $d_{z^2}$ above $E_F$
and $d_{x^2-y^2}$ below $E_F$ along the $N$-$\Gamma$-$X$.
The former provides additional negative contribution
while the latter reduces positive contribution.
As a consequence, $E(\downarrow\downarrow)<0$ for $c/a>1$.
The sign behavior of the $\uparrow\downarrow$-component, $E(\uparrow\downarrow)$,
is completely opposite to $E(\downarrow\downarrow)$,
as all terms in Eq.~(\ref{eq:MCA_}) take opposite signs.\cite{wang93:prb}
For the $\uparrow\uparrow$-component when $c/a<1$, we focus near the $P$-$N$.
The largest positive contribution in the $\downarrow\downarrow$-component
is significantly reduced in the $\uparrow\uparrow$-channel
because the empty $d_{x^2-y^2}$ band in the minority spin
is occupied in the majority spin, and the empty $d_{z^2}$ band contributes negatively.
Thus, $E(\uparrow\uparrow)<0$.
When $c/a>1$, 
the occupancy of $e_g$ states are reversed again due to the level reversal,
therefore $E(\uparrow\uparrow)>0$.
We want to point out that the level reversal between the $d_{x^2-y^2}$ and the $d_{z^2}$ states
not only affects the sign behavior of MCA but also the exchange-splitting in DOS.
Above argument of the sign behavior is more clearly supported
by the {\em k}-resolved MCA analysis. [See SI Fig.~S1-S5.]

In summary, a new metastable phase of the {\em bct}--Ru has been identified 
to exhibit a large PMCA, two orders of magnitude greater than conventional magnetic metals.
In the context of spintronics application,
this large anisotropy along with low magnetization and small volume would be key factors
to realize materials with low switching current and high thermal stability.
Magnetism of the {\em bct}--Ru is mainly governed by the Jahn-Teller split $e_{g}$ states.
As the strength of the tetragonal distortion changes, magnetism of the {\em bct}--Ru
shows an interesting reentrance behavior for $1<c/a<1.1$,
The tetragonal distortion accompanies MCA changes in both magnitudes and signs,
as a result of the level reversal between $d_{x^2-y^2}$ and $d_{z^2}$.

\begin{acknowledgments}
  DO, SHR, and SCH are supported by Basic Research Program (2010−0008842)
  and Priority Research Centers Program (2009−0093818)
  through the National Research Foundation (NRF) funded
  by the Korean Ministry of Education, Science and Technology.
  NP was supported by Basic Science Research Program through
  the NRF funded
  by the Korean Ministry of Education (2013R1A1A2007910).
  KN acknowledges support from a Grant-in-Aid for
  Scientific Research (No. 24540344) from the Japan Society for the Promotion of Science.
  DO and SHR equally contributed authors.
\end{acknowledgments}

%

\end{document}